\documentclass[12pt]{article}
\input{tcilatex}

\begin{document}

\title{Anholonomic Frames, Generalized Killing Equations, and Anisotropic
Taub NUT Spinning Spaces}
\author{Sergiu I. Vacaru \thanks{%
E-mail address:\ vacaru@fisica.ist.utl.pt, ~~ sergiu$_{-}$vacaru@yahoo.com,\
} \\
%EndAName
{\small \textit{Physics Department, CSU Fresno, Fresno, CA 93740--8031, USA,
and }}\\
{\small \textit{Centro Multidisciplinar de Astrofisica - CENTRA,
Departamento de Fisica,}}\\
{\small \textit{Instituto Superior Tecnico, Av. Rovisco Pais 1, Lisboa,
1049-001, Portugal}}\\
{\ }\\
{\small and }\\
{\ }\\
Ovidiu Tintareanu-Mircea \thanks{%
E-mail address:~~ ovidiu@venus.nipne.ro}\\
{\small \textit{Institute of Space Sciences, P.O.Box MG--23, }}\\
{\small \textit{RO 76911, Magurele, Bucharest, Romania}}}
\date{January 31, 2002}
\maketitle

\begin{abstract}
By using anholonomic frames in (pseudo) Riemannian spaces we define
anisotropic extensions of Euclidean Taub--NUT spaces. With respect to
coordinate frames such spaces are described by off-diagonal metrics which
could be diagonalized by corresponding anholonomic transforms. We define the
conditions when the 5D vacuum Einstein equations have as solutions
anisotropic Taub--NUT spaces. The generalized Killing equations for the
configuration space of anisotropically spinning particles (anisotropic
spinning space) are analyzed. Simple solutions of the homogeneous part of
these equations are expressed in terms of some anisotropically modified
Killing-Yano tensors. The general results are applied to the case of the
four-dimensional locally anisotropic Taub-NUT manifold with Euclidean
signature. We emphasize that all constructions are for (pseudo) Riemannian
spaces defined by vacuum soltions, with generic anisotropy, of 5D Einstein
equations, the solutions being generated by applying the moving frame method.

\vskip5pt

Pacs 12.10.-g,12.90.+b, 02.40.+m, 04.20.Me;\

\end{abstract}

%\newpage

%\newpage\setcounter{page}1

\section{Introduction}

Much attention has been paid to off--diagonal metrics in higher dimensional
gravity beginning the Salam, Strathee and Perracci work \cite{sal} which
showed that including off--diagonal components in higher dimensional metrics
is equi\-valent to including $U(1), SU(2)$ and $SU(3)$ gauge fields. The
approach was developed by construction of various locally isotropic
solutions of vacuum 5D Einstein equations describing 4D wormholes and/or
flux tube gravitational--electromagnetic configurations (see Refs. \cite%
{chodos}).

Recently, the off--diagonal metrics were considered in a new fashion by
applying the method of anholonomic frames with associated nonlinear
connections \cite{v,vsbd,vp} which allowed to construct new classes of
solutions of Einstein's equations in three (3D), four (4D) and five (5D)
dimensions which had generic local anisotropy, \textit{e.g.} static black
hole and cosmological solutions with ellipsoidal or toroidal symmetry,
various soliton--dilaton 2D and 3D configurations in 4D gravity, and
wormhole and flux tubes with anisotropic polarizations and/or running
constants with different extensions to backgrounds of rotation ellipsoids,
elliptic cylinders, bipolar and toroidal symmetry and anisotropy.

Another class of 4D metrics induced from 5D Kaluza--Klein theory is
connected with the Euclidean Taub--NUT metric which is involved in many
modern studies of physics, for instance, in definition of the gravitational
analogue of the Yang--Mills instantons \cite{howk} and of Kaluza--Klein
monopole \cite{mant} related with geodesic geometry of higher dimension
(pseudo) Riemannian spaces \cite{atiyah} (see a recent review and original
results in \cite{vv}).

The construction of monopole and instanton solutions, with deformed
symmetries, in modern string theory, extra dimensional gravity and quantum
chromodynamics is of fundamental importance in understanding these theories
(especially their non-perturbative aspects). Such solutions are difficult to
find, and the solutions which are known usually have a high degree of
symmetry. In this work we apply the method of anholonomic frames to
construct the general form anholonomically constrained Taub NUT metrics in
5D Kaluza-Klein theory. These solutions have local anisotropy which would
make their study using holonomic frames difficult. This helps to demonstrate
the usefulness of the anholonomic frames method in studying anisotropic
solutions. Most physical situations do not possess a high degree of
symmetry, and so the anholonomic frames method provides a useful
mathematical framework for studying these less symmetric configurations.

We emphasize that the anholonomic moving frame method works effectively in
construction of anisotropic mass hierarchies with running of constants in
modern brane physics \cite{vbrane} (on new directions in extra dimension
gravity see Refs. \cite{rs}. This allows us to approach a task of of
primordial importance of definition of non--perturbative models and finding
of exact solutions in higher dimension field theory describing anisotropic
monopole/instanton configurations with running constants.

The metrics considered for both wormhole and Taub-NUT geometry and physics
could be given by 5D line elements with 3D spherical coordinates $(r,\theta
,\varphi ),$
\begin{eqnarray}
ds_5^2 &=&-e^{2\nu (r)}dt^2+ds_4^2,  \label{wtnut} \\
ds_4^2 &=&V^{-1}(r)[dr^2+a(r)(d\theta ^2+\sin ^2\theta \,d\varphi ^2)] +
\nonumber \\
& & 16m_0^2 V(r) r_0^2e^{2\psi (r)- 2\nu (r)}[d\chi +\omega (r)dt+n_0\cos
\theta \,d\varphi ]^2  \nonumber
\end{eqnarray}
where the metric coefficients and constants $m_0,r_0,n_0\,$ have to be
correspondingly parametrized in order to select two particular cases:

\begin{enumerate}
\item We must put $a(r)=r^{2},r_{0}^{2}e^{2\psi (r)-2\nu
(r)}=1,n_{0}=1,\omega (r)=0,\nu (r)=0,m_{0}=const,$
\begin{equation}
V(r)=\left( 1+{\frac{4m_{0}}{r}}\right) ^{-1}  \label{vf}
\end{equation}%
and to impose on the fifth coordinate the condition $0\leq \chi <4\pi ,$ $%
4m_{0}(\chi +\varphi )=-x^{5},$ if we want to obtain the Taub-NUT metric
connected with the gauge field $\vec{A}$ of a monopole
\begin{eqnarray}
A_{r}=A_{\varphi } &=&0,~~~A_{\theta }=4m_{0}(1-\cos \theta )  \nonumber \\
\vec{B} &=&rot\vec{A}={\frac{4m_{0}\vec{r}}{r^{3}},}  \label{mfmon}
\end{eqnarray}%
where $\vec{r}$ denotes a three-vector $\vec{r}=(r,\theta ,\varphi );$ the
so called NUT singularity is absent if $x^{5}$ is periodic with period $%
16\pi m_{0}$ \cite{sorkin}.

\item The wormhole / flux tube metrics \cite{chodos} are parametrized if we
put $V(r)=1$ and $16m_0^2=1$ for $r\in \{-R_0,+R_0\}$ $(R_0\leq \infty ),$ $%
r_0=const;$ all functions $\nu (r),\psi (r)$ and $a(r)$ are taken to be even
functions of $r$ satisfying the conditions $\nu ^{\prime }(0)=\psi ^{\prime
}(0)=a^{\prime }(0)=0$. The coefficient $\omega (r)$ is treated as the $t$%
--component of the electromagnetic potential and $n_0\cos \theta $ is the $%
\varphi $-component. These electromagnetic potentials lead to radial
Kaluza-Klein `electrical' $E_{KK}$ and `magnetic'$H_{KK}$ fields:
\begin{equation}
E_{KK}=r_0\omega ^{\prime }e^{3\psi -4\nu }=q_0/a(r)  \label{efw}
\end{equation}
with the `electric' charge $q_0=r_0\omega ^{\prime }(0)$ and
\begin{equation}
H_{KK}=Q_0/a(r)  \label{mfw}
\end{equation}
with `magnetic' charge $Q_0=n_0r_0.$

The solution in \cite{ds} satisfied the boundary conditions $a(0)=1,\psi
(0)=\nu (0)=0$ where it was proved that the free parameters of the metric
are varied there are five classes of wormhole /flux tube solutions.
\end{enumerate}

We note that the metric (\ref{wtnut}) defines solutions of vacuum Einstein
equations only for particular parametrizations of type 1 or 2;\ it is not a
vacuum solution for arbitrary values of coefficients.

The main results of works \cite{vsbd,vp} were obtained by applying the
anholonomic frame method which allowed to construct off--diagonal metrics
describing wormhole / flux tube configurations with anisotropic varying on
the coordinates $\chi $ or $\varphi $ given by some higher dimension
renormalizations of the constants $r_0\rightarrow \widehat{r}_0(...,\chi )$
(or, inversely, $\rightarrow \widehat{r}_0(..,\varphi ))$ and/or $%
n\rightarrow \widehat{n}(...,\chi )$ (or, inversely, $\rightarrow \widehat{n}%
(..,\varphi )).$

The purpose of this paper is to construct Taub-NUT like metrics with
anisotropic variations of the constant $m_0,$ when $m_0\rightarrow
m(...,\chi ),$ or $m_0\rightarrow m(...,\varphi ).$ We note that such
anisotropic metrics are given by off--diagonal coefficients which define
solutions of the 5D vacuum Einstein equations and generalize the
constructions from \cite{sorkin,gross} to Taub--NUT locally anisotropic
gravitational instantons embedded into anisotropic variants of Kaluza--Klein
monopoles (the first anisotropic instanton solutions were proposed in in
Refs. \cite{vg} for the so--called generalized Finsler--Kaluza--Klein spaces
and locally anisotropic gauge gravity, here we note that in this paper we
shall not concern topics relating generalized Lagrange and Finsler (super)
spaces \cite{v});\ by using anholonomic frames we can model anisotropic
instanton configurations in usual Riemannian spaces. The anisotropic metrics
are defined as (pseudo) Riemannian ones which admit a diagonalization with
respect to some anholonomic frame bases with associated nonlinear connection
structures. Such spacetimes, provided with metrics with generic anisotropy
and anholonomic frame structure, are called as anisotropic spaces-time.

Let us introduce a new 5th coordinate
\begin{equation}
\varsigma =\chi - \int \mu (\theta ,\varphi )^{-1} d\xi (\theta ,\varphi )
\label{5new}
\end{equation}
for which\ $d\chi +n_0\cos \theta d\varphi =d\varsigma +n_0\cos \theta
d\theta $\ and
\[
\frac{\partial \xi }{\partial \varphi }=\mu n_0\cos \theta ,\frac{\partial
\xi }{\partial \theta }=-\mu n_0\cos \theta ,
\]
if the factor $\mu (\theta ,\varphi )$ is taken, for instance,
\[
\mu (\theta ,\varphi )= |\cos \theta|^{-1} \exp (\theta -\varphi ),
\mbox{
for } \cos \theta \neq 0.
\]
With respect to the new extradimensional coordinate $\varsigma $ the
component $A_\varphi $ of the electromagnetic potential is removed into the
component $A_\theta ;$ this will allow us to treat the coordinates $%
(t,r,\theta )$ as holonomic coordinates but $\left( \varphi ,\varsigma
\right) $ as anholonomic ones.

For our further considerations it is convenient to use a conformally
transformed (multiplied on the factor $V(r)$) Taub NUT metric with the 5th
coordinate $\varsigma $
\begin{eqnarray}
ds^2 &=&-dt^2+dr^2+r^2(d\theta ^2+\sin ^2\theta d\varphi ^2)  \label{ansatz2}
\\
&&+16m_0^2V^2(r)[d\varsigma +\cos \theta d\theta ]^2  \nonumber
\end{eqnarray}
which will be used for generalizations in order to obtain new solutions of
the vacuum Einstein equations, being anisotropic on coordinates $(\theta
,\varphi ,\varsigma ).$ This metric generates a monopole configuration (\ref%
{mfmon}).

The paper is organized as follow:

Section 2 outlines the geometry of anholonomic frames with associated
nonlinear connections on (pseudo) Riemannian spaces. The metric ansatz for
anisotropic solutions is introduced.

In Section 3, there are analyzed the basic properties of solutions of vacuum
Einstein equations with mixed holonomic and anholonomic variables. The
method of construction of exact solutions with generic local anisotropy is
developed.

In Section 4, we construct three classes of generalized anisotropic Taub NUT
metrics, being solutions of the vacuum Einstein equations, which posses
anisotropies of parameter $m$ on angular coordinate $\varphi ,$ or contains
a running constant $m(\varsigma )$ and/or are elliptically polarized on
angular coordinate $\theta .$

Section 5 is devoted to a new exact 5D vacuum solution for anisotropic Taub
NUT wormholes obtained as a nonlinear superposition of the running on extra
dimension coordinate Taub NUT metric and a background metric describing
locally isotropic wormhole / flux tube configurations.

Section 6 elucidates the problem of definition of integrals of motion for
anholonomic spinning of particles in anisotropic spaces. There are
introduced Killing, energy and momentum and Runge--Lenz vectors with respect
to anholonomic bases with associated nonlinear connection structures defined
by anisotropic solutions of vacuum Einstein equations. There are proposed
and analyzed the action for anisotropic spinning of particles, defined the
Poisson--Dirac brackets on anisotropic spaces. We consider anisotropic
Killing equations and discuss the problem of construction their generic
solutions and non--generic solutions with anholonomic Killing--Yano tensors.

In Section 7, we approach the problem of definition of Killing--Yano tensors
for anisotropic Taub NUT spinning spaces and construct the corresponding Lie
algebra with anisotropic variation of constants.

Finally, in Section 8, some conclusion remarks are presented.

\section{Anholonomic Frames and Nonlinear Connections in Riemannian Spaces}

In this section we outline the basic formulas on anholonomic frames with
mixed holonomic--anholonomic components (variables) and associated nonlinear
connection structures in Riemannian spaces.

\subsection{Metric ansatz}

Let us consider a 5D pseudo--Riemannian spacetime of signature $(-,+,+,+,$ $%
+)$ and denote the local coordinates
\[
u^{\alpha }=(x^{i},y^{a})=(x^{1}=t,x^{2}=r,x^{3}=\theta ,y^{4}=s,y^{5}=p),
\]%
where $\left( s,p\right) =\left( \varsigma ,\varphi \right) ,$ or,
inversely, $\left( s,p\right) =\left( \varphi ,\varsigma \right) $ -- or
more compactly $u=(x,y)$ -- where the Greek indices are conventionally split
into two subsets $x^{i}$ and $y^{a}$ labeled, respectively, by Latin indices
of type $i,j,k,...=1,2,3$ and $a,b,...=4,5.$ The local coordinate bases, $%
\partial _{\alpha }=(\partial _{i},\partial _{a}),$ and their duals, $%
d^{\alpha }=\left( d^{i},d^{a}\right) ,$ are written respectively as
\begin{equation}
\partial _{\alpha }\equiv \frac{\partial }{du^{\alpha }}=(\partial _{i}=%
\frac{\partial }{dx^{i}},\partial _{a}=\frac{\partial }{dy^{a}})
\label{pder}
\end{equation}%
and
\begin{equation}
d^{\alpha }\equiv du^{\alpha }=(d^{i}=dx^{i},d^{a}=dy^{a}).  \label{pdif}
\end{equation}

The 5D (pseudo) Riemannian squared linear interval
\begin{equation}
ds^2=g_{\alpha \beta }du^\alpha du^\beta  \label{metric1}
\end{equation}
is given by the metric coefficients $g_{\alpha \beta }$ (a matrix ansatz
definded with respect to the coordinate frame base (\ref{pdif})) in the form
\begin{equation}
\left[
\begin{array}{ccccc}
-1 & 0 & 0 & 0 & 0 \\
0 & g_2+w_2^{\ 2}h_4+n_2^{\ 2}h_5 & w_2w_3h_4+n_2n_3h_5 & w_2h_4 & n_2h_5 \\
0 & w_3w_2h_4+n_2n_3h_5 & g_3+w_3^{\ 2}h_4+n_3^{\ 2}h_5 & w_3h_4 & n_3h_5 \\
0 & w_2h_4 & w_3h_4 & h_4 & 0 \\
0 & n_2h_5 & n_3h_5 & 0 & h_5%
\end{array}
\right] ,  \label{ansatz0}
\end{equation}
where the coefficients are some necessary smoothly class functions of type:
\begin{eqnarray}
g_{2,3} &=&g_{2,3}(r,\theta )=\exp [2b_{2,3}(r,\theta )],  \label{bvar} \\
h_{4,5} &=&h_{4,5}(r,\theta ,s)=\exp [2f_{4,5}(r,\theta ,s)],  \label{qvar}
\\
w_{2,3} &=&w_{2,3}(r,\theta ,s),n_{2,3}=n_{2,3}(r,\theta ,s);  \nonumber
\end{eqnarray}
one considers dependencies of the coefficients of metric on two so--called
isotropic variables $(r,\theta )$ and on one anisotropic variable $,y^4=s,$
(in similar fashions we can alternatively consider dependencies on arbitrary
couples of $x$--coordinates completed with one $y$--coordinate, for
instance, $(r,\theta )$ and $(r,\theta ,p))$.

The metric (\ref{metric1}) with coefficients (\ref{ansatz0}) can be
equivalently rewritten in the form
\begin{equation}
\delta s^{2}=g_{ij}\left( r,\theta \right) dx^{i}dx^{i}+h_{ab}\left(
r,\theta ,s\right) \delta y^{a}\delta y^{b},  \label{dmetric}
\end{equation}%
with diagonal coefficients
\begin{equation}
g_{ij}=\left[
\begin{array}{lll}
1 & 0 & 0 \\
0 & g_{2} & 0 \\
0 & 0 & g_{3}%
\end{array}%
\right] \mbox{ and }h_{ab}=\left[
\begin{array}{ll}
h_{4} & 0 \\
0 & h_{5}%
\end{array}%
\right]   \label{ansatzd}
\end{equation}%
if instead the coordinate bases (\ref{pder}) and (\ref{pdif}) one introduces
the anholonomic frames (anisotropic bases)
\begin{equation}
{\delta }_{\alpha }\equiv \frac{\delta }{du^{\alpha }}=(\delta _{i}=\partial
_{i}-N_{i}^{b}(u)\ \partial _{b},\partial _{a}=\frac{\partial }{dy^{a}})
\label{dder}
\end{equation}%
and
\begin{equation}
\delta ^{\alpha }\equiv \delta u^{\alpha }=(\delta ^{i}=dx^{i},\delta
^{a}=dy^{a}+N_{k}^{a}(u)\ dx^{k})  \label{ddif}
\end{equation}%
where the $N$--coefficients are parametrized
\[
N_{1}^{4,5}=0,N_{2,3}^{4}=w_{2,3}\mbox{ and }N_{2,3}^{5}=n_{2,3}
\]%
(they define an associated to some anholonomic frames (\ref{dder}) and (\ref%
{ddif}), nonlinear connection, N--connection, structure, see details in Refs %
\cite{ma,v,vst,vsts,vg}). A N--connection induces a global decomposition of
the 5D pseudo--Ri\-e\-man\-nian spacetime into holonomic (horizontal, h) and
anholonomic (vertical, v) directions. In a preliminary form the concept of
N--connections was applied by E. Cartan in his approach to Finsler geometry %
\cite{cartan} and a rigorous definition was given by Barthel \cite{barthel}
(Ref. \cite{ma} gives a modern approach to the geometry of N--connections,
and to generalized Lagrange and Finsler geometry, see also Ref. \cite{vst}
for applications of N--connection formalism in supergravity and superstring
theory). As a particular case one obtains the linear connections if $%
N_{i}^{a}(x,y)=\Gamma _{bi}^{a}\left( x\right) y^{a}.$

A quite surprising result is that the N--connection structures can be
naturally defined on (pseudo) Riemannian spacetimes \cite{v,vst,vg} by
associating them with some anholonomic frame fields (vielbeins) of type (\ref%
{dder}) satisfying the relations\ $\delta _\alpha \delta _\beta -\delta
_\beta \delta _\alpha =W_{\alpha \beta }^\gamma \delta _\gamma , $ with
nontrivial anholonomy coefficients
\begin{eqnarray}
W_{ij}^k &=&0,W_{aj}^k=0,W_{ia}^k=0,W_{ab}^k=,W_{ab}^c=0,  \label{anholonomy}
\\
W_{ij}^a &=&-\Omega _{ij}^a,W_{bj}^a=-\partial _bN_j^a,W_{ia}^b=\partial
_aN_j^b,  \nonumber
\end{eqnarray}
where
\[
\Omega _{ij}^a=\delta _jN_i^a-\delta _iN_j^a
\]
is the nonlinear connection curvature (N--curvature).

One says that the N--connection coefficients model a locally anisotropic
structure on spacetime ( a locally anisotropic spacetime) when the partial
derivative operators and coordinate differentials, (\ref{pder}) and (\ref%
{pdif}), are respectively changed into N--elongated operators (\ref{dder})
and (\ref{ddif}).

Conventionally, the N--coefficients decompose the spa\-ce\-time values
(ten\-sors, spinors and connections) into sets of mixed
holonomic--anholonomic variables (coordinates) provided respectively with
'holonomic' indices of type $i,j,k,...$ and with 'anholonomic' indices of
type $a,b,c,...$. Tensors, metrics and linear connections with coefficients
defined with respect to anholonomic frames (\ref{dder}) and (\ref{ddif}) are
distinguished (d) by N--coefficients into holonomic and anholonomic subsets
and called, in brief, d--tensors, d--metrics and d--connections.

\subsection{D--connections, d--torsions and d--curvatures}

On (pseudo)--Riemannian spacetimes the associated N--connection structure
can be treated as a ''pure'' anholonomic frame effect which is induced if we
are dealing with mixed sets of holonomic--anholonomic basis vectors. When we
are transferring our considerations only to coordinate frames (\ref{pder})
and (\ref{pdif}) the N--connection coefficients are removed into both
off--diagonal and diagonal components of the metric like in (\ref{ansatz0}).
In some cases the N--connection (anholonomic) structure is to be stated in a
non--dynamical form by definition of some initial (boundary) conditions for
the frame structure, following some prescribed symmetries of the
gravitational--matter field interactions, or , in another cases, a subset of
N--coefficients have to be treated as some dynamical variables defined as to
satisfy the Einstein equations.

\subsubsection{D--metrics and d-connections:}

A metric of type (\ref{dmetric}), in general, with arbitrary coefficients $%
g_{ij}\left( x^k,y^a\right) $ and $h_{ab}\left( x^k,y^a\right) $ defined
with respect to a N--elongated basis (\ref{ddif}) is called a d--metric.

A linear connection $D_{\delta _\gamma }\delta _\beta =\Gamma _{\ \beta
\gamma }^\alpha \left( x,y\right) \delta _\alpha ,$ associated to an
operator of covariant derivation $D,$ is compatible with a metric $g_{\alpha
\beta }$ and N--connection structure on a 5D pseudo--Riemannian spacetime if
$D_\alpha g_{\beta \gamma }=0.$ The linear d--connection is parametrized by
irreducible h--v--components,\ $\Gamma _{\ \beta \gamma }^\alpha =\left(
L_{\ jk}^i,L_{\ bk}^a, C_{\ jc}^i,C_{\ bc}^a\right) ,$ where
\begin{eqnarray}
L_{\ jk}^i &=&\frac 12g^{in}\left( \delta _kg_{nj}+\delta _jg_{nk}-\delta
_ng_{jk}\right) ,  \label{dcon} \\
L_{\ bk}^a &=&\partial _bN_k^a+\frac 12h^{ac}\left( \delta
_kh_{bc}-h_{dc}\partial _bN_k^d-h_{db}\partial _cN_k^d\right) ,  \nonumber \\
C_{\ jc}^i &=&\frac 12g^{ik}\partial _cg_{jk},\ C_{\ bc}^a = \frac
12h^{ad}\left( \partial _ch_{db}+\partial _bh_{dc}-\partial _dh_{bc}\right).
\nonumber
\end{eqnarray}
This defines a canonical linear connection (distinguished by a
N--connection) which is similar to the metric connection introduced by
Christoffel symbols in the case of holonomic bases.

\subsubsection{D--torsions and d--curvatures:}

The anholonomic coefficients $w_{\ \alpha \beta }^{\gamma }$ and
N--elongated derivatives give nontrivial coefficients for the torsion
tensor, $T(\delta _{\gamma },\delta _{\beta })=T_{\ \beta \gamma }^{\alpha
}\delta _{\alpha },$ where
\begin{equation}
T_{\ \beta \gamma }^{\alpha }=\Gamma _{\ \beta \gamma }^{\alpha }-\Gamma _{\
\gamma \beta }^{\alpha }+w_{\ \beta \gamma }^{\alpha },  \label{torsion}
\end{equation}%
and for the curvature tensor, $R(\delta _{\tau },\delta _{\gamma })\delta
_{\beta }=R_{\beta \ \gamma \tau }^{\ \alpha }\delta _{\alpha },$ where
\begin{equation}
R_{\beta \ \gamma \tau }^{\ \alpha }=\delta _{\tau }\Gamma _{\ \beta \gamma
}^{\alpha }-\delta _{\gamma }\Gamma _{\ \beta \tau }^{\alpha }+\Gamma _{\
\beta \gamma }^{\varphi }\Gamma _{\ \varphi \tau }^{\alpha }-\Gamma _{\
\beta \tau }^{\varphi }\Gamma _{\ \varphi \gamma }^{\alpha }+\Gamma _{\
\beta \varphi }^{\alpha }w_{\ \gamma \tau }^{\varphi }.  \label{curvature}
\end{equation}%
We emphasize that the torsion tensor on (pseudo) Riemannian spacetimes is
induced by anholonomic frames, whereas its components vanish with respect to
holonomic frames. All tensors are distinguished (d) by the N--connection
structure into irreducible h--v--components, and are called d--tensors. For
instance, the torsion, d--tensor has the following irreducible,
nonvanishing, h--v--components,\ $T_{\ \beta \gamma }^{\alpha }=\{T_{\
jk}^{i},C_{\ ja}^{i},S_{\ bc}^{a},T_{\ ij}^{a},T_{\ bi}^{a}\},$ where
\begin{eqnarray}
T_{.jk}^{i} &=&T_{jk}^{i}=L_{jk}^{i}-L_{kj}^{i},\quad
T_{ja}^{i}=C_{.ja}^{i},\quad T_{aj}^{i}=-C_{ja}^{i},  \nonumber \\
T_{.ja}^{i} &=&0,\quad T_{.bc}^{a}=S_{.bc}^{a}=C_{bc}^{a}-C_{cb}^{a},
\nonumber \\
T_{.ij}^{a} &=&-\Omega _{ij}^{a},\quad T_{.bi}^{a}=\partial
_{b}N_{i}^{a}-L_{.bi}^{a},\quad T_{.ib}^{a}=-T_{.bi}^{a}  \nonumber
\end{eqnarray}%
(the d--torsion is computed by substituting the h--v--components of the
canonical d--connection (\ref{dcon}) and anholonomic coefficients (\ref%
{anholonomy}) into the formula for the torsion coefficients (\ref{torsion}%
)). The curvature d-tensor has the following irreducible, non-vanishing,
h--v--components\ $R_{\beta \ \gamma \tau }^{\ \alpha
}=\{R_{h.jk}^{.i},R_{b.jk}^{.a},P_{j.ka}^{.i},P_{b.ka}^{.c},$ $%
S_{j.bc}^{.i},S_{b.cd}^{.a}\},$\ where
\begin{eqnarray}
R_{h.jk}^{.i} &=&\delta _{k}L_{.hj}^{i}-\delta
_{j}L_{.hk}^{i}+L_{.hj}^{m}L_{mk}^{i}-L_{.hk}^{m}L_{mj}^{i}-C_{.ha}^{i}%
\Omega _{.jk}^{a},  \label{dcurvatures} \\
R_{b.jk}^{.a} &=&\delta _{k}L_{.bj}^{a}-\delta
_{j}L_{.bk}^{a}+L_{.bj}^{c}L_{.ck}^{a}-L_{.bk}^{c}L_{.cj}^{a}-C_{.bc}^{a}%
\Omega _{.jk}^{c},  \nonumber \\
P_{j.ka}^{.i} &=&\partial _{a}L_{.jk}^{i}+C_{.jb}^{i}T_{.ka}^{b}-(\delta
_{k}C_{.ja}^{i}+L_{.lk}^{i}C_{.ja}^{l}-L_{.jk}^{l}C_{.la}^{i}-L_{.ak}^{c}C_{.jc}^{i}),
\nonumber \\
P_{b.ka}^{.c} &=&\partial _{a}L_{.bk}^{c}+C_{.bd}^{c}T_{.ka}^{d}-(\delta
_{k}C_{.ba}^{c}+L_{.dk}^{c\
}C_{.ba}^{d}-L_{.bk}^{d}C_{.da}^{c}-L_{.ak}^{d}C_{.bd}^{c}),  \nonumber \\
S_{j.bc}^{.i} &=&\partial _{c}C_{.jb}^{i}-\partial
_{b}C_{.jc}^{i}+C_{.jb}^{h}C_{.hc}^{i}-C_{.jc}^{h}C_{hb}^{i},  \nonumber \\
S_{b.cd}^{.a} &=&\partial _{d}C_{.bc}^{a}-\partial
_{c}C_{.bd}^{a}+C_{.bc}^{e}C_{.ed}^{a}-C_{.bd}^{e}C_{.ec}^{a}  \nonumber
\end{eqnarray}%
(the d--curvature components are computed in a similar fashion by using the
formula for curvature coefficients (\ref{curvature})).

\section{Einstein Equations with Anholonomic Va\-riables}

In this section we write and analyze the Einstein equations on 5D (pseudo)
Riemannian spacetimes provided with anholonomic frame structures and
associated N--connections.

\subsection{Einstein equations with matter sources}

The Ricci tensor $R_{\beta \gamma }=R_{\beta ~\gamma \alpha }^{~\alpha }$
has the d--components
\begin{eqnarray}
R_{ij} &=&R_{i.jk}^{.k},\quad R_{ia}=-^2P_{ia}=-P_{i.ka}^{.k},
\label{dricci} \\
R_{ai} &=&^1P_{ai}=P_{a.ib}^{.b},\quad R_{ab}=S_{a.bc}^{.c}.  \nonumber
\end{eqnarray}
In general, since $^1P_{ai}\neq ~^2P_{ia}$, the Ricci d-tensor is
non-symmetric (this could be with respect to anholonomic frames of
reference). The scalar curvature of the metric d--connection, $%
\overleftarrow{R}=g^{\beta \gamma }R_{\beta \gamma },$ is computed as
\begin{equation}
{\overleftarrow{R}}=G^{\alpha \beta }R_{\alpha \beta }=\widehat{R}+S,
\label{dscalar}
\end{equation}
where $\widehat{R}=g^{ij}R_{ij}$ and $S=h^{ab}S_{ab}.$

By substituting (\ref{dricci}) and (\ref{dscalar}) into the 5D Einstein
equations
\begin{equation}
R_{\alpha \beta }-\frac 12g_{\alpha \beta }R=\kappa \Upsilon _{\alpha \beta
},  \label{5einstein}
\end{equation}
where $\kappa $ and $\Upsilon _{\alpha \beta }$ are respectively the
coupling constant and the energy--momentum tensor. The definition of matter
sources with respect to anholonomic frames is considered in Refs. \cite{v}.

\subsection{5D vacuum Einstein equations}

In this paper we deal only with vacuum 5D, locally, anisotropic
gravitational equations which in invariant h-- v--components are written
\begin{eqnarray}
R_{ij}-\frac 12\left( \widehat{R}+S\right) g_{ij} &=&0,  \label{einsteq2} \\
S_{ab}-\frac 12\left( \widehat{R}+S\right) h_{ab} &=&0,  \nonumber \\
^1P_{ai} =0,\ ^2P_{ia} &=&0.  \nonumber
\end{eqnarray}

The main `trick' of the anholonomic frames method for integrating the
Einstein equations in general relativity and various (super) string and
higher / lower dimension gravitational theories is to find the coefficients $%
N_j^a$ such that the block matrices $g_{ij}$ and $h_{ab}$ are diagonalized %
\cite{v,vst}. This greatly simplifies computations. With respect to such
anholonomic frames the partial derivatives are N--elongated (locally
anisotropic).

\subsubsection{Non--trivial Ricci components:}

The metric (\ref{metric1}) with coefficients (\ref{ansatz0}) (equivalently,
the d--metric (\ref{dmetric}) with coefficients (\ref{ansatzd})) is assumed
to solve the 5D Einstein vacuum equations $R_{\alpha \beta }=0,$ which are
distinguished in h-- and v--components as
\begin{eqnarray}
R_{2}^{2} &=&R_{3}^{3}=\frac{-1}{2g_{2}g_{3}}\times \lbrack g_{3}^{\bullet
\bullet }-\frac{g_{2}^{\bullet }g_{3}^{\bullet }}{2g_{2}}-\frac{%
(g_{3}^{\bullet })^{2}}{2g_{3}}+g_{2}^{^{\prime \prime }}-\frac{%
g_{2}^{^{\prime }}g_{3}^{^{\prime }}}{2g_{3}}-\frac{(g_{2}^{^{\prime }})^{2}%
}{2g_{2}}]=0,  \label{einsteq3a} \\
&&  \nonumber \\
R_{4}^{4} &=&R_{5}^{5}=-\frac{\beta }{2h_{4}h_{5}}=0,  \label{einsteq3b} \\
&&  \nonumber \\
R_{4i} &=&-w_{i}\frac{\beta }{2h_{5}}-\frac{\alpha _{i}}{2h_{5}}=0,\ i=2,3;
\label{einsteq3c} \\
&&  \nonumber \\
R_{5i} &=&-\frac{h_{5}}{2h_{4}}\left[ n_{i}^{\ast \ast }+\gamma n_{i}^{\ast }%
\right] =0,\ i=2,3;  \label{einsteq3d}
\end{eqnarray}%
where
\begin{eqnarray}
\alpha _{2} &=&{h_{5}^{\ast }}^{\bullet }-\frac{{h_{5}^{\ast }}}{2}\left(
\frac{h_{4}^{\bullet }}{h_{4}}+\frac{h_{5}^{\bullet }}{h_{5}}\right) ,={%
h_{5}^{\ast }}\left( \ln |f_{5}^{\ast }|+f_{5}-f_{4}\right) ^{\bullet },{%
h_{5}^{\ast }\neq 0;}  \label{alpha2} \\
\alpha _{3} &=&{h_{5}^{\ast }}^{\prime }-\frac{{h_{5}^{\ast }}}{2}\left(
\frac{h_{4}^{\prime }}{h_{4}}+\frac{h_{5}^{\prime }}{h_{5}}\right) ,={%
h_{5}^{\ast }}\left( \ln |f_{5}^{\ast }|+f_{5}-f_{4}\right) ^{^{\prime }},{%
h_{5}^{\ast }\neq 0;}  \label{alpha3} \\
\beta &=&h_{5}^{\ast \ast }-\frac{(h_{5}^{\ast })^{2}}{2h_{5}}-\frac{%
h_{5}^{\ast }h_{4}^{\ast }}{2h_{4}},={h_{5}^{\ast }}\left( \ln |f_{5}^{\ast
}|+f_{5}-f_{4}\right) ^{^{\ast }},{h_{5}^{\ast }\neq 0;}  \label{beta} \\
\gamma &=&\frac{3}{2}\frac{h_{5}}{h_{5}}^{\ast }-\frac{h_{4}}{h_{4}}^{\ast }=%
\left[ 3f_{5}-2f_{4}\right] ^{\ast };  \label{gamma}
\end{eqnarray}%
for further applications we gave the formulas with respect to $h_{4,5}$
coefficients of metric as well with respect to $f_{4,5}$, see (\ref{qvar}),
and, for simplicity, the partial derivatives are denoted $h^{\bullet
}=\partial h/\partial x^{2},f^{\prime }=\partial f/\partial x^{3}$ and $%
f^{\ast }=\partial f/\partial s.$

It was possible to construct very general classes of solutions for such
equations \cite{v,vsbd,vp} describing locally an\-isotropic soliton, black
hole, black tori and wormhole objects.

\subsubsection{ General properties of anisotropic vacuum solutions:}

In the vacuum case the equations (\ref{einsteq3a}), (\ref{einsteq3b}), (\ref%
{einsteq3c}) and (\ref{einsteq3d}) form a very simplified system of
equations with separations of variables which can be solved consequently for
every couples of d--metric coefficients, $\left( g_{2},g_{3}\right) ,\left(
h_{4},h_{5}\right) ,$ and N--connection coefficients $w_{2,3}$ and $n_{2,3}$
(see Refs \cite{vbrane} on the main teorems and methods of constructing
exact solutions):

\begin{enumerate}
\item The equation (\ref{einsteq3a}) relates two functions $g_{2}(r,\theta )$
and $g_{3}(r,\theta )$ and their partial derivatives on 'isotropic'
coordinates $r$ and $\theta .$ The solution is trivial if we chose $g_{2}=1$
and $g_{3}=r^{2}$ in order to reduce the coefficients from (\ref{ansatz0}),
respectively, to those from (\ref{ansatz2}).

\item The equation (\ref{einsteq3b}) contains partial derivatives only on
anisotropic coordinate $s$ and relates two functions $h_{4}(r,\theta ,s)$
and $h_{5}(r,\theta ,s).$ This equation is satisfied by arbitrary two
functions $h_{4}(r,\theta ,s)$ and $h_{5}(r,\theta )$ for which $h_{5}^{\ast
}=0.$

If the condition $h_{5}^{\ast }\neq 0$ is satisfied, we can write (\ref%
{einsteq3b}), in $f$--variables (see (\ref{bvar})), as
\[
\left( \ln |f_{5}^{\ast }|+f_{5}-f_{4}\right) ^{\ast }=0,
\]%
which is solved by arbitrary functions $f_{5}(r,\theta ,s)$ and
\begin{equation}
f_{4}=f_{4[0]}+\ln |f_{5}^{\ast }|+f_{5},  \label{sole3b}
\end{equation}%
where $f_{4[0]}=f_{4[0]}(r,\theta ).$ The general solution of (\ref%
{einsteq3b}) expressing $h_{5}$ via $h_{4}$ is
\begin{eqnarray}
h_{5} &=&[h_{5[1]}(r,\theta )+h_{5[2]}(r,\theta )\int \sqrt{h_{4}(r,\theta
,s)}ds]^{2},  \nonumber \\
&=&h_{5[0]}(r,\theta )[1+\varpi (r,\theta )s]^{2},h_{4}^{\ast }=0,
\label{zeroht}
\end{eqnarray}%
for some functions $f_{5[0,1,2]}(r,\theta )$ and $\varpi (r,\theta )$\
stated by boundary conditions and locally anisotropic limits as well from
the conditions that the equations (\ref{einsteq3c}) and (\ref{einsteq3d})
are compatible. Inversely, for a prescribed value of $h_{5},$ the general
solution of (\ref{einsteq3b}) is (\ref{sole3b}) which can be rewritten with
respect to variables $h_{4,5},$
\begin{eqnarray}
h_{4} &=&h_{4[0]}(r,\theta )\left[ \left( \sqrt{|h_{5}(r,\theta ,s)|}\right)
^{\ast }\right] ^{2}  \label{sole3b1} \\
&=&h_{4}(r,\theta ,s),\mbox{ an arbitrary function if }h_{5}^{\ast }=0,
\nonumber
\end{eqnarray}

\item If the functions $h_{4}(r,\theta ,s)$ and $h_{5}(r,\theta ,s)$ were
defined, the equations (\ref{einsteq3c}) can be solved as independent linear
algebraic equations for $w_{2,3},$\ $w_{\widehat{i}}\beta +\alpha _{\widehat{%
i}}=0,\widehat{i}=2,3.$ \ For zero matter sources this is a trivial result
because in this case the conditions $\beta =0$ and $\alpha _{\widehat{i}}=0$
(see the formulas ($\ref{alpha2}$), ($\ref{alpha3}$) and ($\ref{beta}$)) are
automatically fulfilled. In consequence, the resulting sourceless equations (%
\ref{einsteq3c}) became some trivial equations admitting arbitrary values of
functions $w_{\widehat{i}}\left( r,\theta ,s\right) ;$ such functions can be
associated to some coordinate transforms for vanishing anholonomy
coefficients $W_{\alpha \beta }^{4}=0,$ see (\ref{anholonomy}), or to some
anholonomy coefficients, in such cases being not contained in the vacuum
Einstein equations, which must be stated by some boundary and symmetry
conditions.

\item The equations (\ref{einsteq3c}) can be solved in general form if the
functions $h_{4}(r,\theta ,s)$ and $h_{5}(r,\theta ,s)$ (and, in
consequence, the coefficient $\gamma $ from (\ref{gamma})) are known,
\begin{eqnarray}
n_{2,3}(r,\theta ,s)=n_{2,3[0]}(r,\theta ) &{+}&n_{2,3[1]}(r,\theta )\int
\frac{h_{4}(r,\theta ,s)}{{h}_{5}^{3/2}(r,\theta ,s)}ds,\ \gamma \neq 0;
\nonumber  \label{ncoef} \\
n_{2,3}(r,\theta ,s) &=&n_{2,3[0]}(r,\theta )+n_{2,3[1]}(r,\theta )s,\gamma
=0,  \nonumber
\end{eqnarray}%
where the functions $n_{2,3[0]}(r,\theta )$ and $n_{2,3[1]}(r,\theta )$
should be defined from some boun\-da\-ry conditions.
\end{enumerate}

\section{Taub NUT Metrics with Anisotropies and Running of Constant}

The conformally transformed Taub NUT metric (\ref{ansatz2}) can be
considered as a locally isotropic background with trivial vanishing local
anisotropies. By coordinate transforms of the 5th coordinate and a conformal
transform on two holonomic coordinates (here we mention that in two
dimensions the coordinate and conformal transforms are equivalent) the
isotropic background metric can be transformed into a form parameterizing
the usual Taub NUT solution of the vacuum Einstein equations.

The aim of this section is to construct and analyze three types of
anisotro\-pic generalizations of the Taub NUT solution.

The data parameterizing a metric (\ref{ansatz0}) (equivalently, a d--metric (%
\ref{dmetric}), or (\ref{ansatz2})) are
\begin{eqnarray}
x^{1} &=&t,x^{2}=r,x^{3}=\theta ,\ y^{4}=s=\varsigma ,y^{5}=p=\varphi ,
\nonumber \\
g_{1} &=&-1,g_{2}=1,g_{3}=r^{2},h_{4}=r^{2}\sin ^{2}\theta ,\
h_{5}=16m_{0}^{2}V^{2}(r),  \nonumber \\
w_{i} &=&0,n_{1}=0,n_{2}=0,n_{3}=\cos \theta  \label{set1}
\end{eqnarray}%
which defines a trivial (locally isotropic) solution of the vacuum Einstein
equations (\ref{einsteq3a})--(\ref{einsteq3d}) satisfying the conditions $%
h_{4,5}^{\ast }=0.$

\subsection{Generalizations of Taub NUT solutions to anisot\-ro\-pies and
running of constant}

The simplest way to obtain anisotropic Taub NUT like solutions, is to follow
the approach developed for generation anisotropic black hole \cite{v,vg} and
wormhole / flux tube solutions \cite{vsbd} when the constants like mass and
charges are considered to be effectively anisotropically polarized by some
anholonomic (anisotropic) higher dimension interactions;\ in our case is to
consider that the parameter $m_0$ from (\ref{ansatz2}) (see also (\ref{set1}%
)) is not a constant but a renormalized value $\,m_0\to m=m(r,\theta ,s).$

\subsubsection{Taub NUT\ metrics with anisotropic\newline
running of constants on $s=\protect\varsigma $}

We generate from the isotropic solution (\ref{set1}) a new one, anisotropic,
in this manner: Let us consider the case when\ $h_{4}(r,\theta )=r^{2}\sin
^{2}\theta $ when $h_{4}^{\ast }=0,$ but $h_{5}^{\ast }\neq 0.$ Following
the solution (\ref{zeroht}) we parameterize
\begin{equation}
16m^{2}(r,\theta ,\varsigma )=16m_{0}^{2}[1+\varpi (r,\theta )\varsigma ]^{2}
\label{slin}
\end{equation}%
and
\[
h_{5}(r,\theta ,\varsigma )=16m_{0}^{2}V^{2}(r)[1+\varpi (r,\theta
)\varsigma ]^{2},
\]%
where $\widehat{V}(r,\theta ,\varsigma )$ is just the function (\ref{vf})
but defined by a renormalized value $m(r,\theta ,\varsigma ).$

The conditions of vanishing of constants (\ref{alpha2}) and (\ref{alpha3})
are also satisfied if $\varpi (\varsigma ,\theta )=\varpi _{0}=const.$ For
simplicity, in this subsection, we shall consider solutions with ''pure''
running of the constant $m$ on $\varsigma $ and the function
\begin{equation}
\widehat{V}(r,\varsigma )=\left( 1+{\frac{4m(\varsigma )}{r}}\right) ^{-1}.
\label{vf1}
\end{equation}

Because (in this case) the coefficients $\beta $ and $\alpha _{2,3},$ and in
consequence. $w_{2,3}$ could be arbitrary functions (see formulas (\ref%
{alpha2}), (\ref{alpha3}) and (\ref{beta}) and equation (\ref{einsteq3c}));
in the locally isotropic limit, $\varpi \chi \rightarrow 0,$ we put that $%
w_{2,3}\rightarrow 0$. The values $n_{2,3}$ depends on anisotropic variable $%
\varsigma $ as follow from the solution (\ref{ncoef}) with $\gamma =3\varpi
_0.$

We obtain the locally isotropic limit (\ref{set1}), for $\varpi \varsigma
\rightarrow 0,$ if we fix the boundary conditions with $%
n_{2[0,1]}=0,n_{3[0]}=0$ but $n_{3[1]}(r,\theta )=\cos \theta .$

So, a parametrization of the ansatz (\ref{ansatz0}),
\begin{eqnarray}
\delta s^{2} &=&-dt^{2}+dr^{2}+r^{2}d\theta ^{2}+r^{2}\sin ^{2}\theta
d\varsigma ^{2}  \label{set2} \\
&&+16m_{0}^{2}[1+\varpi (r,\theta )\varsigma ]^{2}{\widehat{V}}%
^{2}(r,\varsigma )\left[ d\varphi +\cos \theta \exp (3\varpi _{0}\varsigma
)d\theta \right] ^{2},  \nonumber
\end{eqnarray}
defines a locally anisotropic solution of the vacuum Einstein equations (\ref%
{einsteq3a})--(\ref{einsteq3d}) generalizing the Taub NUT solution (\ref%
{ansatz2}). We can treat the solution (\ref{set2}) as describing an
anisotropic Kaluza--Klein monopole with running constant (on extra dimension
coordinate) obtained by embedding the anisotropic Taub NUT gravitational
instanton into 5D theory, adding the coordinate in a way as to be compatible
with running of constant of effective magnetic configurations (in brief, we
shall call such solutions as $\varsigma $--solutions).

We conclude that the solutions describing gravitational monopoles and
instantons \cite{sorkin,gross} can be generalized to some anisotropic
configurations with running constants.

\subsubsection{ Taub NUT\ metric with anisotropy \newline
of constants on angle $\protect\varphi $}

In a similar fashion we can consider anisotropic (angular) dependencies of
constants with $s=\varphi $ (in brief, we call such solutions as $\varphi $%
--solutions). The simplest way is to take $h_{5}^{\ast }=0$ but $h_{4}^{\ast
}\neq 0,$ i.e to define a solution with
\begin{eqnarray*}
h_{4}(r,\theta ,\varphi ) &=&16m_{0}^{2}\exp [\varpi (r,\theta ,\varphi )]{%
\widehat{V}}^{2}(r,\theta ,\varphi ), \\
h_{5}(r,\theta ) &=&r^{2}\sin ^{2}\theta ,
\end{eqnarray*}%
where $\widehat{V}(r,\theta ,\varphi )$ is the function (\ref{vf}) but
defined by the renormalized parameter $m(r,\theta ,\varphi )$%
\begin{equation}
\widehat{V}(r,\theta ,\varphi )=\left( 1+{\frac{4m(r,\theta ,\varphi )}{r}}%
\right) ^{-1};  \label{vf1a}
\end{equation}%
we can take arbitrary values of $w_{2,3}$ because $\beta =0$ and $\alpha _{%
\widehat{i}}$ from the equations (\ref{einsteq3c}) vanish.

For small, constant, polarizations we can approximate
\[
h_{4}(r,\theta ,\varphi )=16m_{0}^{2}{V}^{2}(r)[1+\varpi \varphi ]
\]%
and consider only anisotropic angular variations of the constant, $m\sim
m(\varphi ).$

The d--metric
\begin{eqnarray}
\delta s^{2} = -dt^{2}+ dr^{2}+r^{2}d\theta ^{2}+16m_{0}^{2}\exp
[\varpi (r,\theta ,\varphi )]{\widehat{V}}^{2}(r,\theta ,\varphi
)(d\varphi +\cos \theta d\theta )^{2} \label{set2b}  \\
+ r^{2}\sin ^{2}\theta \left[ d\varsigma +n_{2[1]}(r,\theta
)dr\int \varpi (r,\theta ,\varphi )d\varphi + n_{3[1]}(r,\theta
)d\theta \int \varpi (r,\theta ,\varphi )\right] ^{2},  \nonumber
\end{eqnarray}
models a locally anisotropic generalization of the solution (\ref{ansatz2})
for anisotropic dependencies of the constant $m$ on angle $\varphi $ which
describe a 5D Kaluza Klein monopole with angular anisotropic constant
obtained by embedding the anisotropic Taub NUT gravitational instanton into
5D theory.

So, in this subsection, we constructed two classes of generalized,
anisotro\-pic, Taub NUT solutions of the 5D vacuum Einstein equations:\ the
first class is for $s=\varphi $ (\textit{i.e.} anisotropic polarizations)
and the second is for $s=\varsigma $ (\textit{i.e.}with dependence of the
constant $m$ on the fifth coordinate). The metric (\ref{ansatz0}),
describing these two classes of solutions, can be written with respect to a
coordinate frame (\ref{pder}) where the existence of non--diagonal terms is
emphasized.

\subsection{Gravitational $\protect\theta $--elliptic polarizations of the
Taub NUT solutions}

This subsection is devoted to such generalizations of the Taub NUT\ metrics
when the anisotropies of constant are modeled by elliptic polarizations on
the angle $\theta ,$ i. e. the constant is renormalized as $m_0\rightarrow
m(\theta ,\varsigma )$ $($or $m(\theta ,\varphi ))\sim m_1(1+\varepsilon
_r\cos \theta )^{-1}$ where $\varepsilon _r$ is the eccentricity of an
ellipse and $m_1=m_1(\varsigma )$ (or $m_1=m_1(\varphi ))$ is defined as for
the solution (\ref{set2}), in brief, for $\varsigma $--solutions (or as for
the solution (\ref{set2b}), in brief, for $\varphi $--solution). Because the
method of construction of such solutions is very similar to that considered
in the previous subsection, we shall omit the details on integration of
equations and present only the final values of the d--metric and
N--connection coefficients. The results can be verified by straightforward
calculations.

\subsubsection{Elliptic polarizations for $\protect\varsigma $--solutions}

The simplest way of definition of such polarizations for $\varsigma $%
--solutions (\ref{zeroht}) and (\ref{slin}) is to consider the off--diagonal
metric
\begin{eqnarray}
\delta s^{2} &=&-dt^{2}+dr^{2}+r^{2}d\theta ^{2}+r^{2}\sin ^{2}\theta
d\varsigma ^{2} +16m_{0}^{2}\left[ 1+\varepsilon _{r}\cos \theta \right]
^{-2} \times  \label{set3a} \\
&&[1+\varpi (r,\theta )\varsigma ]^{2}\ {\widehat{V}}^{2}(r,\theta
,\varsigma )\left[ d\varphi +\cos \theta \exp (3\varpi _{0}\varsigma
)d\theta \right] ^{2},  \nonumber
\end{eqnarray}
the function $\widehat{V}(r,\theta ,\varsigma )$ is the function (\ref{vf})
redefined by an ellipsoidally renormalized coefficient $m\sim m_{0}\left[
1+\varepsilon _{r}\cos \theta \right] ^{-1},$%
\[
\widehat{V}(r,\theta ,\varsigma )=\left( 1+{\frac{4m_{0}[1+\varpi (r,\theta
)\varsigma ]}{r(1+\varepsilon _{r}\cos \theta )}}\right) ^{-1}
\]%
where there is also a linear dependence on extradimension coordinate $%
\varsigma ;$\ for simplicity, we can choose $\varpi (r,\theta )=\varpi
_{0}=const.$

\subsubsection{Elliptic polarizations for $\protect\varphi $--solutions}

Such $\varphi $--solutions with elliptic variations of the constant $m$ on
angle $\theta $ are distinguished by the metric
\begin{eqnarray}
& &\delta s^{2} =-dt^{2}+dr^{2}+r^{2}d\theta ^{2}  \nonumber \\
&& +16m_{0}^{2}\left[ 1+\varepsilon _{r}\cos \theta \right] ^{-2}\exp
[\varpi (r,\theta ,\varphi )]{\widehat{V}}^{2}(r,\theta ,\varphi )(d\varphi
+\cos \theta d\theta )^{2} +  \nonumber \\
&& r^{2}\sin ^{2}\theta \left[ d\varsigma +n_{2[1]}(r,\theta )dr\int \varpi
(r,\theta ,\varphi )d\varphi +n_{3[1]}(r,\theta )d\theta \int \varpi
(r,\theta ,\varphi )\right] ^{2},  \label{set3b}
\end{eqnarray}
where $\widehat{V}(r,\theta ,\varphi )$ generalize the functions (\ref{vf1}%
), (\ref{vf1a}) for anisotropic dependencies like
\[
m(r,\theta ,\varphi )\sim m_{1}(\varphi )\left[ 1+\varepsilon _{r}\cos
\theta \right] ^{-1},
\]%
\[
{\widehat{V}}(r,\theta ,\varphi )=\left( 1+{\frac{4m_{1}(\varphi )}{r\left[
1+\varepsilon _{r}\cos \theta \right] }}\right) ^{-1}.
\]%
The constructed solutions of 5D vacuum Einstein equations, (\ref{set3a}) and
(\ref{set3b}) contain some additional elliptic polarizations comparing with
respective solutions (\ref{set2}) and (\ref{set2b}). Such 5D Kaluza--Klein
monopoles induced from 4D Taub NUT instantons behave theirselves as some
objects with running on $\varsigma ,$ or anisotropic on $\varphi ,$ constant
$m$ which is also elliptically polarized on the angle $\theta .$

\section{A Superposition of Wormhole and Taub NUT Metrics}

By applying the method of anholonomic frames we can construct nonlinear
anisotropic superpositions of the Taub NUT metric with some metrics defining
wormhole / flux tube configurations:\ this way one defines anisotropic
generalizations of the metric (\ref{wtnut}) which are called \textbf{%
anisotropic Taub NUT wormholes.} For simplicity, in this paper, we consider
only a $\varsigma $--solution for the Taub NUT configurations with the
wormhole background chosen as to be locally isotropic (see Refs \cite{ds},
on isotropic wormholes and flux tubes, and \cite{vsbd}, on locally
anisotropic wormhole solutions).

For our purpose the ansatz (\ref{ansatz0}) is generalized by introducing the
coefficient $N_1^6=n_1(r)=\omega (r)$ of the nonlinear connection and, for
simplicity, we consider $w_i=0$ and $n_2=0.$
\begin{equation}
\left[
\begin{array}{ccccc}
-1+n_1^{\ 2}h_5 & 0 & n_1n_3h_5 & 0 & n_1h_5 \\
0 & g_2 & 0 & 0 & 0 \\
n_1n_3h_5 & 0 & g_3+n_3^{\ 2}h_5 & 0 & n_3h_5 \\
0 & 0 & 0 & h_4 & 0 \\
n_1h_5 & 0 & n_3h_5 & 0 & h_5%
\end{array}
\right] ,  \label{ansatz3}
\end{equation}
which results in a diagonal d--metric
\begin{eqnarray*}
\delta s^2 &=&-dt^2+g_2(r,\theta )dr^2+g_3(r,\theta )d\theta ^2  \nonumber \\
& &+h_4(r,\theta,\varsigma )\delta \varsigma ^2+h_5(r,\theta ,\varsigma )
\delta \varphi ^2  \nonumber \\
\delta \varsigma &=&d\varsigma ,\delta \varphi =d\varphi
+n_1(r)dt+(1+n_0)\cos \theta d\theta  \nonumber
\end{eqnarray*}
where the coefficients $n_1(r)=\omega (r)$ and $n_3(\theta)=(1+n_0)\cos
\theta $ and the constant $n_0$ are introduced for the $t$ and $\theta $
components of the wormhole electromagnetic potential like in (\ref{wtnut}).

The data defining a vacuum $\varsigma $--solution for the (\ref{ansatz3}),
including the solution (\ref{set2}) into a wormhole background are given by
the coefficients of the d--metric
\begin{eqnarray}
\delta s^{2} &=&-dt^{2}+dr^{2}+a(r)d\theta ^{2}+a(r)\sin ^{2}\theta
d\varsigma ^{2}+16m_{0}^{2}r_{0}^{2}\exp [2\psi (r)]\times  \label{nutw} \\
&&[1+\varpi (r,\theta )\varsigma ]^{2}{\widehat{V}}^{2}(r,\varsigma )\left[
d\varphi +\omega (r)dr+(1+n_{0})\cos \theta \exp (3\varpi _{0}\varsigma
)d\theta \right] ^{2},  \nonumber
\end{eqnarray}
where $\widehat{V}(r,\varsigma )$ and $V(r)$ are respectively those from (%
\ref{vf1}) and (\ref{vf}) and the constant $r_{0}^{2}$ and term $\exp [2\psi
(r)]$ were introduced as some multiples defining the wormhole / flux tube
configuration.

The proprieties of the Taub NUT monopole with running constant $m(\varsigma
) $ are the same as those stated for the solution (\ref{set2}), resulting in
a similar to (\ref{mfmon}) magnetic field components
\begin{eqnarray}
A_r^{(nut)}=A_\varphi ^{(nut)} &=&0,~~~A_\theta ^{(nut)}=4m(\varsigma
)(1-\cos \theta )  \nonumber \\
\vec{B}^{(nut)} &=&rot\vec{A}^{(nut)}={\frac{4m(\varsigma )\vec{r}}{r^3}.}
\label{mfmon1}
\end{eqnarray}
embedded into a static background of wormhole / flux tube configurations
defined by additional components of electric, $E_{KK}^{(worm)}=q_0/a(r),$
and magnetic, $H_{KK}^{(worm)}=Q_0/a(r),$ fields as it is defined
respectively by the formulas (\ref{efw}) and (\ref{mfw}).

As the free parameters of the wormhole background (see Ref \cite{ds}) are
varied there are five classes of solutions with the properties:

\begin{enumerate}
\item $Q=0$ or $H_{KK}^{(worm)}=0$, a wormhole--like `electric' object;

\item $q=0$ or $E_{KK}^{(worm)}=0$, a finite `magnetic' flux tube;

\item $H_{KK}^{(worm)}=E_{KK}^{(worm)}$, an infinite `electromagnetic' flux
tube;

\item $H_{KK}^{(worm)}<E_{KK}^{(worm)}$, a wormhole--like `electromagnetic'
object;

\item $H_{KK}^{(worm)}>E_{KK}^{(worm)}$, a finite, `magnetic--electric' flux
tube.
\end{enumerate}

The number wormhole / flux tube classes must be extended to new
configurations which arise in the presence of the magnetic field of the Taub
NUT monopole, for instance, by considering the 6th class for the pure
monopole configuration; we must take into account possible contribution of
the monopole magnetic field to the structure of magnetic flux tubes,
wormhole--like electromagnetic object and/or magnetic--electric flux tubes
by analyzing the total magnetic field $H_{KK}^{(worm)}+B^{(nut)}$ with
possible (elliptic and another type) vacuum gravitational polarizations.

Finally, in this section, we remark that in a similar fashion we can
construct $\varphi $--solutions with an anisotropic parameter $m(\varphi )$
describing an anisot\-ro\-pic Taub NUT monopole embedded into a wormhole /
flux tube background and to generalized both $\varsigma $-- and $\varphi $%
--solutions for various configurations with elliptic polarizations and
rotation hypersurface symmetries (like rotation ellipsoids, elliptic
cylinders, bipolars and tori) as we constructed exact solutions of 3D-5D
Einstein equations describing anisotropic black holes/tori and anisotropic
wormholes \cite{v,vsbd,vp}.

\section{Anisotropically Spinning Particles and Integrals of Motion}

In the previous two sections we proved that the Taub NUT and wormhole
metrics admit various type of anisotropic generalizations modeled by
anholonomic frames with associated N--connection structure. It is of
interest the investigation of symmetries of such an\-iso\-tro\-pic spaces
and definition of corresponding invariants of motion of spinning particles.
The general rules for developing of corresponding geodesic calculus and
definition of generalized anisotropic Killing vectors are:

\begin{itemize}
\item We have to change the partial derivatives and differentials into
N--elongated ones (\ref{pder}) and (\ref{pdif}) by redefinition of usual
formulas for developing a formalism of differential, integral and
variational calculus on anisotropic spaces.

\item The metric, linear connection, curvature and Ricci tensors have to be
changed into respective d--objects (d--metric (\ref{dmetric}), d--connection
(\ref{dcon}) and corresponding curvature and Ricci d--tensors); the
d--torsion on (pseudo) Riemannian spaces should be treated as an anholonomic
frame effect which vanishes with respect to coordinate bases.

\item The Greek indices $\alpha ,\beta ,...$ should be split into horizontal
$(i,$ $j,$ $...)$ and vertical $(a,b,...)$ ones which on necessity will
point to some holonomic--anholonomic character of variables.

\item By using d--metrics and d--connections the dif\-fe\-ren\-ti\-al,
in\-teg\-ral and va\-ri\-a\-ti\-on\-al calculus on Ri\-e\-man\-ni\-an
manifolds is adapted to the anholonomic frame structure with associated
N--connection. With respect to N--adapted frames (\ref{pder}) and (\ref{pdif}%
) the geometry is similar to the usual Riemannian one when the anholonomic
(constrained dynamics) is coded into the coefficients of N--connection
modeling a local anisotropy.

\item As a matter of principle, all constructions defined with respect to
anholonomic bases can be removed with respect to usual coordinate fra\-mes,
but in this case the metrics became generically off--diagonal and a number
of symmetries (and their constraints) of manifolds are hidden in rather
sophisticate structures and relations for redefined holonomic objects.
\end{itemize}

\subsection{Moving of spinless particles in anisotropic Taub NUT spaces}

\subsubsection{Killing d--vectors:}

The geodesic motion of a spinless particle of unit mass moving into a
background stated by a d--metric $g_{\alpha \beta }=\left(
g_{ij},h_{ab}\right),$ see (\ref{dmetric})) can be derived from the action:
\begin{eqnarray*}
{S} &=&{\frac 12}\int_a^bd\tau \,\,g_{\mu \nu }(u)\,\frac{du}{d\tau }^\mu \,%
\frac{du}{d\tau }^\nu \\
&=&{\frac 12}\int_a^b d\tau\ [g_{ij}(x,y)\frac{dx}{d\tau }^i\,\frac{dx}{%
d\tau }^j + h_{ab}(x,y)\,\frac{dy}{d\tau }^a\, \frac{dy}{d\tau }^a] ,
\end{eqnarray*}
where $\tau $ is a parameter.

The invariance of d--metrics defining anisotropic generalizations of Taub
NUT metrics under spatial rotations and $\varsigma $ translations is
generated by four Killing d--vectors which are obtained by anholonomic
transforms of the usual Killing vectors into corresponding one with
elongated partial derivatives $\partial _\mu \rightarrow \delta _\mu =
\left( \partial _i,\delta _a\right) ,$\ where the partial derivative on the
new 5th coordinate, are $\frac \partial {\partial \varsigma }=\frac{\partial
\chi }{\partial \varsigma }\frac \partial {\partial \chi },$ see (\ref{5new}%
). We write the Killing d--vectors as
\[
D^{(\alpha )} =R^{(\alpha )\mu }\,\delta _\mu , =R^{(\alpha )i}\partial
_i+R^{(\alpha )a}\delta _a, ~~~~\alpha =1,\cdots ,4
\]
or, in details,
\begin{eqnarray*}
D^{(1)} &=&-\sin \varphi \,{\frac \delta {\partial \theta }}-\cos \varphi
\,\cot \theta \,{\frac \partial {\partial \varphi }}+{\frac{\cos \varphi }{%
\sin \theta }}\,\frac{\partial \varsigma }{\partial \chi }{\frac \partial
{\partial \varsigma }}, \\
D^{(2)} &=&\cos \varphi \,{\frac \delta {\partial \theta }}-\sin \varphi
\,\cot \theta \,{\frac \partial {\partial \varphi }}+{\frac{\sin \varphi }{%
\sin \theta }}\frac{\partial \varsigma }{\partial \chi }{\frac \partial
{\partial \varsigma }}, \\
D^{(3)} &=&{\frac \partial {\partial \varphi }},\ D^{(4)} =\frac{\partial
\varsigma }{\partial \chi }{\frac \partial {\partial \varsigma }},
\end{eqnarray*}
following general considerations we give the formulas for arbitrary
transforms of the 5th coordinate, $\varsigma \rightarrow \varsigma \left(
\chi ,....\right) ,$ and elongations of derivatives of type $\delta
/\partial \theta =\partial /\partial \theta -n_3\partial /\partial \varsigma
.$

\subsubsection{Energy and momentum d--vector:}

We know that in the purely locally isotropic bosonic case such invariances
correspond to conservation of angular momentum and ''relative electric
charge'' \cite{vv,17ch}. For anisotropic Taub NUTS we can define similar
objects by using anholonomic transforms of the values given with respect to
coordinate bases into the corresponding values with an\-isot\-ro\-pic
coefficients and variables given with respect to anholonomic frames:
\begin{eqnarray*}
\vec{j} &=&\vec{r}\times \vec{p}\,+\,q\,{\frac{\vec{r}}{r},\ \mbox{ for }%
s=\varsigma (\mbox{or}s=\varphi ),} \\
q &=&16m^{2}\left( r,\theta ,s\right) \,V^{2}(r,\theta ,s)\,(\frac{\partial
\chi }{\partial \varsigma }\frac{d\varsigma }{d\tau }+\cos \theta \,\frac{%
d\theta }{d\tau })
\end{eqnarray*}%
where $\vec{p}=\,d\vec{r}/{d\tau }$ is the ''mechanical momentum'' which is
only part of the momentum canonically conjugate to $\vec{r}$ (in
anisotropization of the momentum formula we do not write the multiple $%
V^{-1} $ because the d--metric used for anisotropic constructions has its
locally isotropic limit being multiplied on the $V$ as a conformal factor,
see (\ref{ansatz2})). The energy is defined by
\begin{eqnarray*}
E &=&{\frac{1}{2}}\,g^{\mu \nu }\,\Pi _{\mu }\,\Pi _{\nu }\,={\frac{1}{2}}%
g^{ij}\,\Pi _{i}\,\Pi _{j}+{\frac{1}{2}}h^{ab}\,\Pi _{a}\,\Pi _{b}\, \\
&=&{\frac{1}{2}}\left[ \vec{r}^{\,2}+\left( {\frac{q^{2}(r,\theta ,s)}{%
4m^{2}(r,\theta ,s)}}\right) ^{2}\right]
\end{eqnarray*}%
is also conserved , $\Pi _{\mu }=\left( \Pi _{i},\Pi _{a}\right) $ being the
covariant momentum d--vector; this value variates with respect to coordinate
frames but behaves itselve as a usual energy with respect to N--adapted
frames.

\subsubsection{Runge--Lenz d--vector:}

There is a conserved vector analogous to the Runge- Lenz vector of the
Coulomb problem in the locally isotropic case \cite{vv,17ch,19rh}, which
with respect to anholonomic frames with anisotropic variables in spite of
the fact of complexity of anholonomic motion in the anisotropic Taub- NUT
spaces defined in previous Sections,
\[
\vec{{K}}=\vec{K}_{\mu \nu }\frac{du^{\mu }}{d\tau }\frac{du^{\nu }}{d\tau }=%
\vec{p}\times \vec{j}+\,\left( {\frac{q^{2}}{4m}}-4mE\right) {\frac{\vec{r}}{%
r}}
\]%
which implies that the trajectories are anisotropic deformations of conic
sections.

\subsection{Spinning of particles with respect to anholonomic frames}

The pseudo-classical limit of the Dirac theory of a spin 1/2 fermion in
curved spacetime is described by the supersymmetric extension of the usual
relativistic point-particle \cite{1r} (the theory of spinors on spaces with
generic local anisotropy was developed in Refs. \cite{vsts}, see also some
models of locally anisotropic supergravity and superstring theories in Refs. %
\cite{vst}).

In this work, the configuration space of spinning particles in anisotropic
space(anisotropic spinning space) is an extension of an ordinary Riemannian
manifold provided with an anholonomic frame and associated N--connection
structure, parametrized by local coordinates $\left\{ u^\mu
=(x^i,y^a)\right\} $, to a graded manifold parametrized by local coordinates
$\left\{ u^\mu ,\psi ^\mu \right\} $, with the first set of variables being
Grassmann-even (commuting) and the second set Grassmann-odd
(anti-commuting). We emphasize that in anholonomic spaces distinguished by a
N--connection structure we must define spinor and Grassmann variables
separately on h--subspace (with holonomic variables) and on v--subspace
(with anholonomic variables).

\subsubsection{Action for anisotropically spinning particles:}

The equation of motion of an anisotropically spinning particle on a
autoparallel (geodesic) is derived from the action:
\begin{eqnarray*}
S =\int d\tau \left( \frac{1}{2}g_{\mu \nu }(u)\frac{du^{\mu }}{d\tau }\frac{%
du^{\nu }}{d\tau }+\frac{i}{2}g_{\mu \nu }(u)\psi ^{\mu }\frac{D\psi ^{\nu }%
}{D\tau }\right) = \\
\frac{1}{2}\int d\tau (g_{ij}(u)\frac{dx^{i}}{d\tau }\frac{dx^{j}}{d\tau }%
+ig_{ij}(u)\psi ^{i}\frac{D\psi ^{j}}{D\tau }+h_{ab}(u)\frac{dy^{a}}{d\tau }%
\frac{dy^{b}}{d\tau }+ih_{ab}(u)\psi ^{a}\frac{D\psi ^{b}}{D\tau
})
\end{eqnarray*}%
where $i^{2}=-1.$

The corresponding world-line anholonomic Ha\-mil\-to\-ni\-an is given by:
\[
H=\frac 12g^{\mu \nu }\Pi _\mu \Pi _\nu =\frac 12(g^{ij}\Pi _i\Pi
_j+h^{ab}\Pi _a\Pi _b)
\]
where $\Pi _\mu =g_{\mu \nu }\frac{du^\mu }{d\tau }= \{g_{ij}\frac{dx^j}{%
d\tau },h_{ab}\frac{dy^b}{d\tau }\}$ is the covariant momentum d--vector.

\subsubsection{Poisson--Dirac brackets:}

For any integral (we use the term integral instead of the usual one,
constant, because on anisotropic spaces we can define any conservations lows
with respect to N--adapted anholonomic frames; the invariants of such
conservations laws are not constant with respect to coordinate frames) of
anholonomic motion ${J}(u,\Pi ,\psi )$, the bracket with $H$ vanishes, $%
\left\{ H,{J}\right\} =0,$\ where the Poisson-Dirac brackets for functions
of the covariant phase-space variables $(u,\Pi ,\psi )$ is defined
\begin{eqnarray*}
\left\{ F,G\right\} &=&{D}_{\mu }F\frac{\partial G}{\partial \Pi _{\mu }}-%
\frac{\partial F}{\partial \Pi _{\mu }}{D}_{\mu }G-{R}_{\mu \nu }\frac{%
\partial F}{\partial \Pi _{\mu }}\frac{\partial G}{\partial \Pi _{\nu }}%
+i(-1)^{a_{F}}\frac{\partial F}{\partial \psi ^{\mu }}\frac{\partial G}{%
\partial \psi _{\mu }}; \\
&=&{D}_{i}F\frac{\partial G}{\partial \Pi _{i}}-\frac{\partial F}{\partial
\Pi _{i}}{D}_{i}G-{R}_{ij}\frac{\partial F}{\partial \Pi _{i}}\frac{\partial
G}{\partial \Pi _{j}}+i(-1)^{a_{F}}\frac{\partial F}{\partial \psi ^{i}}%
\frac{\partial G}{\partial \psi _{i}} \\
&& +{D}_{a}F\frac{\partial G}{\partial \Pi _{a}} -\frac{\partial F}{\partial
\Pi _{a}}{D}_{a}G-{R}_{ab}\frac{\partial F}{\partial \Pi _{a}}\frac{\partial
G}{\partial \Pi _{b}}+i(-1)^{a_{F}}\frac{\partial F}{\partial \psi ^{a}}%
\frac{\partial G}{\partial \psi _{a}}.
\end{eqnarray*}%
In definition of $\{...\}$ there are used the operators
\begin{eqnarray}
{D}_{\mu }F &=&\partial _{\mu }+\Gamma _{\mu \nu }^{\lambda }\Pi _{\lambda }%
\frac{\partial F}{\partial \Pi _{\nu }}-\Gamma _{\mu \nu }^{\lambda }\psi
^{\nu }\frac{\partial F}{\partial \psi ^{\lambda }},  \nonumber \\
{R}_{\mu \nu } &=&\frac{i}{2}\psi ^{\rho }\psi ^{\sigma }R_{\rho \sigma \mu
\nu },  \nonumber
\end{eqnarray}%
where on anisotropic spaces $\Gamma _{\mu \nu }^{\lambda }$ is the canonical
d--connection (\ref{dcon}), $R_{\rho \sigma \mu \nu }$ is the curvature
d--tensor (\ref{curvature}) with components (\ref{dcurvatures}) and $a_{F}$
is the Grassmann parity of $F$ : $a_{F}=(0,1)$ for $F$=(even,odd).

\subsubsection{Anisotropic Killing equations:}

Expanding ${J}(u,\Pi ,\psi )$ in a power series on the canonical momentum,\ $%
{J}=\sum_{n=0}^{\infty }\frac{1}{n!}{J}^{(n)\mu _{1}\dots \mu _{n}}(u,\psi
)\Pi _{\mu _{1}}\dots \Pi _{\mu _{n}}$\ we conclude that the bracket $\{H,{J}%
\}$ vanishes for arbitrary $\Pi _{\mu }$ if and only if the components of ${J%
}$ satisfy the generalized anisotropic Killing equations \cite{1r} :
\begin{equation}
{J}_{(\mu _{1}\dots \mu _{n};\mu _{n+1})}^{(n)}+\frac{\partial {J}_{(\mu
_{1}\dots \mu _{n}}^{(n)}}{\partial \psi ^{\sigma }}\Gamma _{\mu
_{n+1})\lambda }^{\sigma }\psi ^{\lambda }=\frac{i}{2}\psi ^{\rho }\psi
^{\sigma }R_{\rho \sigma \nu (\mu _{n+1}}{{J}^{(n+1)\nu }}_{\mu _{1}\dots
\mu _{n})}  \label{killing6}
\end{equation}%
where the round brackets $(....)$ denote full symmetrization over the
indices enclosed and the covariant derivation '';'' is defined by the
canonical d--connection (\ref{dcon}) and one should be emphasized that every
Greek index split into horizontal and vertical groups like $\mu _{n}=\left(
i_{n},a_{n}\right) $ which results that this equation will contain both
''pure'' horizontal or vertical terms as well terms of ''mixed'' character,
like ${J}_{i_{1}\dots a_{n}}^{(n)}.$

The type of solutions of the generalized anisotropic Killing equations (\ref%
{killing6}) is defined by two classes from the locally isotropic limit \cite%
{2rh,3grh}: the first class of solutions are \textit{generic} ones, which
exists for any spinning particle model and the second class of solutions are
\textit{non-generic} ones, which depend on the specific background space and
anisotropy considered.

\subsubsection{Generic solutions of Killing equations:}

The proper-time translations and supersymmetry are generated by the
Ha\-mil\-to\-ni\-an and supercharge
\begin{equation}
Q_0=\Pi _\mu \psi ^\mu =\Pi _i\psi ^i+\Pi _a\psi ^a  \label{scharge}
\end{equation}
and belong to the first class. There is also an additional ''chiral''
symmetry generated by the chiral charge
\[
\Gamma _{*}=\frac{i^{[\frac d2]}}{d!}\sqrt{g}\epsilon _{\mu _1\dots \mu
_d}\psi ^{\mu _1}\dots \psi ^{\mu _d}
\]
and a dual supersymmetry with generator
\[
Q^{*}=i\{\Gamma _{*},Q_0\}=\frac{i^{[\frac d2]}}{(d-1)!}\sqrt{g}\epsilon
_{\mu _1\dots \mu _d}\Pi ^{\mu _1}\psi ^{\mu _2}\dots \psi ^{\mu _d}
\]
where $d$ is the dimension of spacetime.

\subsubsection{Non--generic solutions and Killing--Yano d--tensors:}

The \textit{non-generic} conserved quantities depend on the explicit form of
the metric $g_{\mu \nu }(u)$ and, in our case, on N--connection structure.
Following Ref. \cite{3grh}, generalizing the constructions for anisotropic
spaces, we introduce the Killing-Yano d--tensors as objects generating
\textit{non-generic} N--distinguished supersymmetries. A d--tensor $f_{\mu
_1\dots \mu _r}$ is called Killing-Yano of valence $r$ if it is totally
antisymmetric and satisfies the equation\ $f_{\mu _1\dots \mu _{r-1}(\mu
_r;\lambda )}=0.$

The method of solution of the system of coupled differential equations (\ref%
{killing6}) is similar to the method developed for locally isotropic spaces %
\cite{vv}, that why here we present only the key results which have to be
split on h-- and v-- indices if we need explicit formulas for
holonomic--anholonomic components.

We start with a $\tilde{{J}}_{\mu _1\dots \mu _n}^{(n)}$ solution of the
homogeneous equation:
\[
\tilde{{J}}_{(\mu _1\dots \mu _n;\mu _{n+1})}^{(n)}+\frac{\partial \tilde{{J}%
}_{(\mu _1\dots \mu _n}^{(n)}}{\partial \psi ^\sigma }\Gamma _{\mu
_{n+1})\lambda }^\sigma \psi ^\lambda =0.
\]

This solution is introduced in the r.h.s. of (\ref{killing6}) for ${J}_{\mu
_1\dots \mu _{n-1}}^{(n-1)}$ and the iteration is carried on to $n=0$.

For the bosonic case the first equation shows that ${J}_0$ is a trivial
constant, the next one is the equation for the Killing d--vectors and so on.
In general, the homogeneous equation for a given $n$ defines a Killing
d--tensor ${J}_{\mu _1\dots \mu _n}^{(n)}$ for which ${J}_{\mu _1\dots \mu
_n}^{(n)}\Pi ^{\mu _1}\dots \Pi ^{\mu _n} $ is a first integral of the
geodesic equation \cite{4dr}. This does not hold for the spinning particles.

Let us consider the case $n=0$, when\ $\tilde{{J}}^{(0)}=\frac i4f_{\mu \nu
}\psi ^\mu \psi ^\nu $ is a solution if $f_{\mu \nu }$ is a Killing-Yano
d--tensor --covariantly constant, i. e. $\tilde{{J}}^{(0)}$ is a separately
conserved quantity. For $n=1$, the natural solution is:\ $\tilde{{J}}_\mu
^{(1)}=R_\mu f_{\lambda \sigma }\psi ^\lambda \psi ^\sigma, $ where $R_\mu $
is a Killing d--vector ($R_{(\mu ;\nu )}=0$) and $f_{\lambda \sigma }$ is a
Killing-Yano d--tensor d--covariantly constant. Introducing this solution in
the r.h.s. of the equation (\ref{killing6}) with $n=0$, we get\ ${J}%
^{(0)}=\frac i2R_{[\mu ;\nu ]}f_{\lambda \sigma }\psi ^\mu \psi ^\nu \psi
^\lambda \psi ^\sigma, $ where the square bracket denotes the
antisymmetrization with norm one.

We define a new integral of anholonomic motion which is peculiar to the
spinning case and has its analogous in the locally isotropic limit:
\[
{J}=f_{\mu \nu }\psi ^\mu \psi ^\nu \left( R_\lambda \Pi ^\lambda +\frac
i2R_{[\lambda ;\sigma ]}\psi ^\lambda \psi ^\sigma \right) .
\]

We can generate another $\psi $-dependent solution of the $n=1$ by starting
from a Killing-Yano d--tensor with $r$ indices,\ $\tilde{{J}}_{\mu
_{1}}^{(1)}=f_{\mu _{1}\mu _{2}\dots \mu _{r}}\psi ^{\mu _{2}}\dots \psi
^{\mu _{r}},$ or, following the above prescription, we express
\[
{J}^{(0)}=\frac{i}{r+1}(-1)^{r+1}f_{[\mu _{1}\dots \mu _{r};\mu _{r+1}]}\psi
^{\mu _{1}}\dots \psi ^{\mu _{r+1}}
\]%
stating that the integral of motion corresponding to these solutions of the
Killing equations is:
\begin{equation}
Q_{f}=f_{\mu _{1}\dots \mu _{r}}\Pi ^{\mu _{1}}\psi ^{\mu _{2}}\dots \psi
^{\mu _{r}}+\frac{i}{r+1}(-1)^{r+1}f_{[\mu _{1}\dots \mu _{r};\mu
_{r+1}]}\psi ^{\mu _{1}}\dots \psi ^{\mu _{r+1}}.  \nonumber
\end{equation}

We conclude that the existence of a Killing-Yano d--tensor with $r$ indices
is equivalent to the existence of a supersymmetry for the spinning space
with supercharge $Q_f$ which anticommutes with $Q_0$, such constructions are
anholonomic and distinguished by the N--connection structure.

\section{Anisotropic Taub-NUT spinning space}

There are four Killing-Yano tensors in the usual, locally isotropic,
Taub-NUT geometry \cite{9gr} which for anisotropic spaces are transformed
into corresponding d--tensors for an\-isot\-ro\-pic Taub NUT spaces,
\[
f_{i}=8m(r,\theta ,s)\delta \varsigma \wedge \delta x_{i}-\epsilon _{ijk}(1+%
\frac{4m(r,\theta ,s)}{r})\delta x_{j}\wedge \delta x_{k}
\]%
which are d--covariantly constant and the fourth Killing-Yano d--tensor is
%\begin{eqnarray}
\[
f_{Y} =8m(r,\theta ,s)\delta \varsigma \wedge dr  +4r(r+2m(r,\theta ,s))(1+%
\frac{r}{4m(r,\theta ,s)})\sin \theta d\theta \wedge \delta \varphi \nonumber
\]
%\end{eqnarray}%
having only one non-vanishing component of the field strength\ ${f_{Y}}%
_{r\theta ;\varphi }=2(1+r/{4m(r,\theta ,s)})r\sin \theta ,$ where $\delta
\varsigma =d\varsigma +\cos \theta d\theta ,$ and the values $\delta x_{j}$
are N--elongated for the v--components.

The corresponding supercharges constructed from the Killing-Yano d-tensors
are $Q_i$ and $Q_Y$. The supercharges $Q_i$ together with $Q_0$ from (\ref%
{scharge}) realize the N=4 supersymmetry algebra \cite{14h}, in our case
distinguished by the N--connection structure\cite{vst} :
\[
\left\{ Q_A,Q_B\right\} =-2i\delta _{AB}H~~~,~~~A,B=0,\dots ,3
\]
making manifest the link between the existence of the Killing-Yano
d--tensors and the hyper-K\"{a}hler d--geometry of the anisotropic Taub-NUT
manifold. Such distinguished manifolds and geometries are constructed as the
usual ones but with respect to N--connection decompositions on
holonomic--an\-ho\-lo\-no\-mic variables.

Starting with these results from the bosonic sector of the Taub-NUT space
one can proceed with the spin contributions. The first generalized Killing
equation (\ref{killing6}) shows that with each Killing vector $R_A^\mu $
there is an associated Killing scalar $B_A.$ The expression for the Killing
scalar is taken as in Ref.\cite{14h}:
\[
B_A=\frac i2R_{A[\mu ;\nu ]}\psi ^\mu \psi ^\nu
\]
with that modification that we use a d--covariant derivation which gives
that the total angular momentum and ''relative electric charge'' become in
the anisotropic spinning case
\[
{\vec{J}}={\vec{B}}+{\vec{j},}J_0=B_0+q
\]
where ${\vec{J}}=(J_1,J_2,J_3)$ and ${\vec{B}}=(B_1,B_2,B_3)$ are given with
respect to anholonomic bases. These integrals of motion are superinvariant: $%
\left\{ J_A,Q_0\right\} =0; A=0,\dots ,3.$

We can introduce a Lie algebra with anholonomic constraints defined by the
Killing d--vectors and realized by the integral of motion through the
Poisson-Dirac brackets.

Now, introducing the Killing d--tensors ${\vec{K}}_{\mu \nu }$ into the
generalized anisot\-ro\-pic Killing equation (\ref{killing6}) we obtain that
the corresponding Killing d--vectors ${{\vec{R}}}_\mu $ have a spin
dependent part ${\vec{S}}_\mu $ as in the locally isotropic case \cite{vv}, $%
{\vec{R}}_\mu =\vec{R}_\mu +\vec{S}_\mu ,$ where $\vec{R}_\mu $ are the
standard Killing d--vectors. The $\psi $-dependent parts of the Killing
d--vectors $\vec{S}_\mu $ contributes to the Runge-Lenz d--vector for the
anisotropic spinning space
\[
\vec{{K}}=\frac 12\vec{K}_{\mu \nu }\cdot \frac{du^\mu }{d\tau }\frac{du^\nu
}{d\tau }+\vec{S}_\mu \cdot \frac{du^\mu }{d\tau },
\]
or in terms of the supercharges $Q_{\underline{i}}$ and $Q_Y$,
\[
\vec{{K}}_i=i\left\{ Q_Y,Q_{\underline{i}}\right\} ~~~,~~~\underline{{i}}%
=1,2,3
\]
we are using underlined Latin indices like $\underline{{i}},\underline{j}%
.,...$ enumerating the number of supercharges, in order to distinguish them
from the h--indices $i,j,...$ used for holonomic variables on anisotropic
spacetime.

The non-vanishing Poisson brackets are (after some algebra):
\begin{eqnarray*}
\left\{ J_{\underline{i}},J\underline{_j}\right\} &=& \epsilon _{\underline{i%
}\underline{j}\underline{k}}J_{\underline{k}},\left\{ J_{\underline{i}},{K}_{%
\underline{j}}\right\} =\epsilon _{\underline{i}\underline{j}\underline{k}}{K%
}_{\underline{k}} \\
\left\{ {K}_{\underline{i}},{K}_{\underline{j}}\right\} &=&\left( \frac{J_0^2%
}{16m^2}-2E\right) \epsilon _{\underline{i}\underline{j}\underline{k}}J_{%
\underline{k}}
\end{eqnarray*}
where the values have anisotropic dependencies. Following the fact of
existence of the Killing-Yano covariantly constants d--tensors ${f_{%
\underline{i}}}_{\mu \nu }$, we can define three integrals of motion:\ $%
S_i=\frac i4f_{\underline{i}\mu \nu }\psi ^\mu \psi ^\nu, \underline{i}%
=1,2,3, $ which realize an $SO(3)$ Lie--algebra similar to that of the
angular momentum:\ $\left\{ S\underline{_i},S_{\underline{j}}\right\} =
\epsilon _{\underline{i}\underline{j}\underline{k}}S_{\underline{k}}.$

These components of the spin are separately conserved, do not depend on the
frame of reference, holonomic or anholonomic, and can be combined with the
angular momentum $\vec{J}$ in order to get a new improved form of the
angular momentum $I_{\underline{i}}=J_{\underline{i}}-S_{\underline{i}}$
with the property that one preserves the algebra\ $\left\{ I_{\underline{i}%
},I_{\underline{j}}\right\} =\epsilon _{\underline{i}\underline{j}\underline{%
k}}I_{\underline{k}}$\ and that it commutes with the $SO(3)$ algebra
generated by the spin $S_{\underline{i}},$ $\left\{ I_{\underline{i}},S_{%
\underline{j}}\right\} =0. $

Let us consider the Dirac brackets of $S_i$ with supercharges\ $\left\{ S_{%
\underline{i}},Q_0\right\} =-\frac{Q_{\underline{i}}}2;~\left\{ S_{%
\underline{i}},Q\underline{_j}\right\} =\frac 12(\delta _{\underline{i}%
\underline{j}}Q_0+\epsilon _{\underline{i}\underline{j}\underline{k}}O_{%
\underline{k}}).$\ We can combine the above presented two $SO(3)$ algebras
to obtain the generators of a conserved $SO(4)$ symmetry among the constants
of motion, a standard basis for which is spanned by $M_{\underline{i}}^{\pm
}=I_{\underline{i}}\pm S_{\underline{i}}.$

There are also possible the combinations \cite{vv}:
\begin{eqnarray*}
\tilde{{J}}_{A\underline{j}\mu }^{(1)}&=&R_{A\mu }f_{\underline{j}\lambda
\sigma }\psi ^\lambda \psi ^\sigma ,~~~A=0,\dots ,3; \\
{J}_{A\underline{j}} &=&f_{\underline{j}_\lambda \sigma }\psi ^\lambda \psi
^\sigma \left( R_{A\mu }\Pi ^\mu +\frac i2R_{A[\alpha ;\beta ]}\psi ^\alpha
\psi ^\beta \right) \\
&=&-4iS_{\underline{j}}J_A~~~,~~~A=0,\dots ,3;\underline{j}=1,2,3.
\end{eqnarray*}
So, there are a sort of Runge-Lenz d--vectors involving only Grassmann
components:
\[
L_{\underline{i}}=\frac 1m\epsilon _{\underline{i}\underline{j}\underline{k}%
}S_{\underline{j}}J_{\underline{k}}~~~;~~~\underline{i},\underline{j},%
\underline{k}=1,2,3;
\]
with the commutation relations:
\[
\{L_{\underline{i}},J_{\underline{j}}\} =\epsilon _{\underline{i}\underline{j%
}\underline{k}}J_{\underline{k}},\ \{L_{\underline{i}},L_{\underline{j}}\} =
\left( \vec{S}\vec{J}-\vec{S}^2\right) \frac 1{m^2}\epsilon _{\underline{i}
\underline{j}\underline{k}}J_{\underline{k}}.
\]

Finally, we note the following Dirac brackets of $L_{\underline{i}}$ with
supercharges:
\begin{eqnarray*}
\{L_{\underline{i}},Q_0\} &=&-\frac 1{2m}\epsilon _{\underline{i}\underline{j%
}\underline{k}}J_{\underline{k}}Q_{\underline{j}} \\
\{L_{\underline{i}},Q_{\underline{j}}\} &=&\frac 1{2m}\left( \epsilon _{%
\underline{i}\underline{j}\underline{k}}Q_0J_{\underline{k}}-\delta _{%
\underline{i}\underline{j}}Q_{\underline{k}}M_{\underline{k}}^{-}+Q_{%
\underline{i}}M_{\underline{j}}^{-}\right) ,
\end{eqnarray*}
and emphasize that the presented algebraic relations hold true for
anisotropic (as on some parameters) dependencies of the constant $m=m\left(
r,\theta ,s\right) $ because we are working with respect to anholonomic
frames locally adapted to the N--connection structure.

\section{Conclusion Remarks}

In this paper, we have extended the method of construction of new exact
solutions, with generic local anisotropy, of the Einstein equations by using
anholonomic frames with associated nonlinear connection structure (the
method was proposed and developed in Refs. \cite{v,vst,vsts,vg}) in order to
generate vacuum metrics defining locally anisotropic Taub NUT instanton and
Kaluza Klein monopoles. Such metrics are off--diagonal with respect to usual
coordinate bases and reflect possible interactions of gravitational fields
with gauge fields, induced from higher dimension gravity, in a new fashion
when the constants interactions could run on the 5th dimension and/or
polarized to some anisotropic configurations on angular coordinates.

The first key result of this paper is the proof that the introduced ansatz
for the metric and anholonomy coefficients substantially simplifies the
structure of resulting vacuum gravitational field equations, the variables
being separated and the system of nonlinear partial equations admitting
exact solutions. In consequence, a straightforward definition of new classes
of anisotropic Taub NUT metrics with the effective constant $m$ varying on
the 5th coordinate, as well with anisotropies and elliptic polarizations on
angular coordinates, was possible. There were emphasized classes of
anisotropic Taub NUT wormhole solutions which can be generalized to various
type of rotation hypersurface backgrounds and deformations.

The second key result is connected with the definition of integrals of
motion of scalar and spinning particles in curved spacetimes provided with
anholonomic frame structure. We proved that the symmetries of such
generalized anisotro\-pic Taub NUT spaces are connected with anholonomic
Killing vectors and tensors which are subjected to some anholonomic
conservation laws. The problem of generation of non--generic anisotropic
supersymmetries was solved by introducing Killing--Yano tensors adapted to
the anholonomic spacetime structure.

Finally, we note that the results of this paper were extended for  3D
solitonic configurations \cite{vp}.

\subsection*{Acknowledgements}

~~ The authors thank M. Visinescu for hospitality, substantial support and
discussion of results. S. V. is grateful to D. Singleton, E. Gaburov and D.
Gontsa for collaboration and discussing of results. He also thanks V. Manu
for support and help. The work is supported both by "The 2000--2001
California State University Legislative Award" and a NATO/Portugal
fellowship for the Instituto Superior Tecnico, Lisabon.

%\end{references}

\end{document}